\documentclass[aps,prb,twocolumn,superscriptaddress,amsmath,amssymb,floatfix]{revtex4-1}

\usepackage{graphicx}
\usepackage[usenames]{color} 
\usepackage{comment} 
\usepackage{bm}
\usepackage{amsmath}

\newcounter{abc}

\newcommand{\br}{{\bf r}} 
\newcommand{\dbr}{{\rm d}{\bf r}}

\newcommand{\bu}{{\bf u}} 
 
\newcommand{\rmin}{r_{\rm min}} 

\newcommand{\be}{\begin{equation}} 
\newcommand{\ee}{\end{equation}}
\newcommand{\bea}{\begin{eqnarray}} 
\newcommand{\eea}{\end{eqnarray}}

\begin{document}

\title{Three-body interactions in complex fluids: virial coefficients from simulation finite-size effects}
\author{Douglas J. Ashton}  
\author{Nigel B. Wilding} 
\affiliation{Department of Physics, University of Bath, Bath BA2 7AY,
United Kingdom} 

\begin{abstract}

A simulation technique is described for quantifying the contribution
of three-body interactions to the thermodynamical properties of
coarse-grained representations of complex fluids. The method is based
on comparing the third virial coefficient $B_3$ for a complex fluid
with that of an approximate coarse-grained model described by a pair
potential. To obtain $B_3$ we introduce a new technique which
expresses its value in terms of the measured volume-dependent
asymptote of a certain structural function. The strategy is applicable
to both Molecular Dynamics and Monte Carlo simulation. Its utility is
illustrated via measurements of three-body effects in models of star
polymer and highly size-asymmetrical colloid-polymer mixtures.

\end{abstract}

\maketitle

\section{Introduction}

\label{sec:intro}

The task of determining the thermodynamical properties of complex fluids
by analytical or computational means is often complicated by a
profusion of degrees of freedom on small length scales. For instance
in order to simulate a system of large flexible molecules such as
polymers or biomolecules, considerable computational effort must be
invested to deal with the vibrational motion of the individual
atoms. Since such motion typically occurs on much shorter timescales
than the relaxation of the system as a whole, this causes difficulty
in probing thermodynamical behaviour. Similarly for systems such as
colloidal dispersions, in which large colloid particles are immersed in
a sea of much smaller particles, the relaxation of the large particles
is typically extremely slow. This is because the small particles hem
in the large ones, hindering their motion.

To tackle such problems ``coarse-graining'' strategies have been
developed. These seek to integrate out the degrees of freedom on short
length scales, leaving a simpler system in which the surviving
coordinates are assumed to interact via effective interactions. In
principle if the coarse-graining is exact, the effective interactions
should account exactly for the effects of the degrees of freedom which
have been subsumed. Often, however, the task of performing an exact
coarse-graining is extremely challenging, chiefly because the
effective potential is many-body in character even when the underlying
interactions are pairwise additive. It is therefore common practice to
implement an approximate coarse-graining in which the full many-body
effective potential is replaced by a simpler one involving only pair
interactions. Such an approximation is widely used in theories and
simulations of the complex fluids because the pair potential itself is
usually straightforward to obtain, either analytically or from a
simulation of two molecules. Once it is obtained one can use it to
study the properties of an $N$ particle system
\cite{Dijkstra1999,Louis:2000mi,Largo2006}.

\begin{figure}[h]
\includegraphics[type=pdf,ext=.pdf,read=.pdf,width=0.95\columnwidth,clip=true]{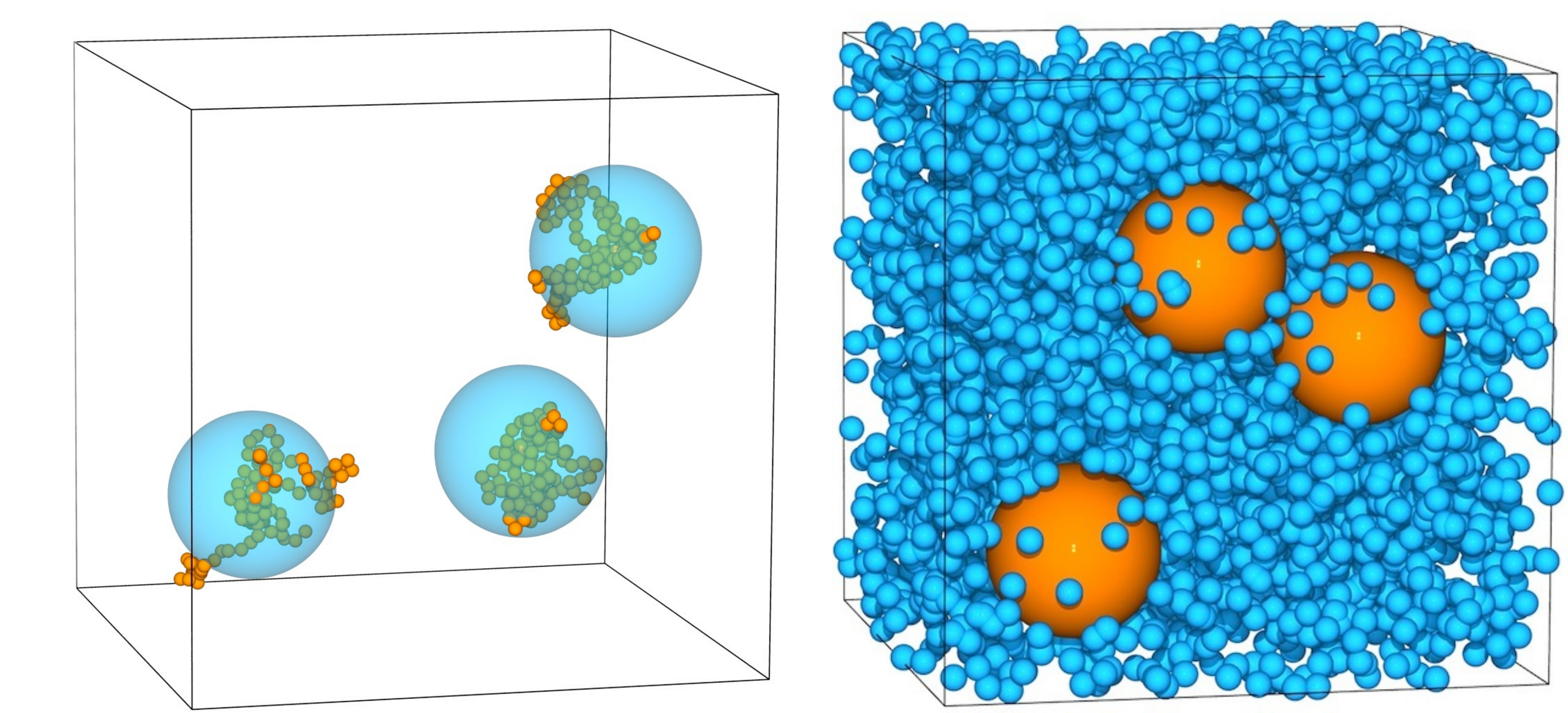}

\caption{ (Left part) A snapshot of three star polymers. The big spheres
  represent a coarse grained model in which each polymer is replaced
  by a single effective particle described by a potential of mean
  force. (Right part) Snapshot of a highly size-asymmetrical mixture of
  spheres. The effective one-component model is realized by tracing
  over the small sphere degrees of freedom.}

\label{fig:cg}
\end{figure}

Two examples of systems to which coarse-graining is frequently applied
are displayed in Fig.~\ref{fig:cg}. In the first, a system of star
polymers, each molecule is replaced by a single soft effective
particle centred on the core atom. These particles interact via a soft
pair potential (the ``potential of mean force'') reflecting the fact
that two star polymers can substantially overlap. In the second
example, a highly size-asymmetrical binary mixture, the small spheres
mediate interactions between the large ones known as ``depletion''
forces \cite{Lekkerkerker:2011}.  Formally the effective interaction
between the large particles is many-body in form, but this is
typically approximated in terms of a depletion pair potential.

Given the prevalence of the pair potential approximation in
coarse-grained representations of complex fluids, there is a need to
be able to quantify its effects on the thermodynamics of model
systems.  In this paper, we describe a method for determining the
scale of three-body interactions which will usually be the most
prominent of the neglected many-body terms in the approximate
coarse-grained model \footnote{A short report on this work has
  appeared elsewhere \protect\cite{Ashton:2014fk}}. Our approach is
based on a comparison of the third virial coefficient $B_3$ of the
full model with that of its pair potential representation. To obtain
$B_3$ in both cases, we introduce a new technique which relates its
value to the finite-size effects in the asymptote of a
simple-to-measure structural quantity.

The organization of the paper as follows. In Sec.~\ref{sec:virials} we
briefly review existing methods for calculating virial coefficients
and discuss why they seem (as yet) unequal to the task of dealing with complex
fluids. Sec.~\ref{sec:method} describes the statistical mechanical
background to our method for determining low order virial
coefficients. A commentary regarding optimization issues, and the
wider context of the method is given in Sec.~\ref{sec:commentary}.
Thereafter in Secs.~\ref{sec:mixture} and \ref{sec:stars} we apply it
to quantify the contribution of three-body interactions to the
third virial coefficient for two coarse-grained models of complex
fluids. A summary of our findings features in Sec.~\ref{sec:discuss}.

\section{Routes to virial coefficients} 
\label{sec:virials}

The virial coefficients appear in the virial expansion of the equation
of state of many-particle systems and as such are key quantities in
the thermodynamical description of fluids. Their utility is manifold. At
a basic level they measure the deviation of the fluid properties from
those of an ideal gas.  More generally, they provide a framework for
systematically calculating the thermodynamical properties of a fluid and
relating these to the nature of the microscopic interactions,
including features such as potential range~\cite{Shaul:2010zr},
molecular flexibility~\cite{DAdamo:2012ys,Shaul:2011fk} and many-body
forces~\cite{Hellmann:2011oq}.

For these and other reasons, substantial activity has been devoted to
calculating virial coefficients for model systems.  Much of this
effort has focused on prototypical fluid models such as hard and soft
spheres and spheroids \cite{Masters:2008qc}. Almost
invariably, the approach taken involves a direct assault on the
cluster integrals that provide a compact mathematical representation
of the virial coefficients in terms of Mayer functions
\cite{Hansen-MacDonald}. We shall refer to this as the Mayer function
route. For low order virials it is often possible to make analytical
progress via this route for simple models. For higher
order virials and molecular systems, numerical simulation is generally
required~\cite{Pratt:1982eq,Labik:2003vf}. In this latter context a
particularly powerful technique is the Mayer Sampling Method (MSM)
  \cite{Singh:2004zr,Schultz:2009fu,Kim:2013ff}. This Monte Carlo (MC) scheme
employs ideas borrowed from free energy perturbation methods to relate
the virial coefficients of the system of interest to those of a
reference system for which the virial coefficients are independently
known. For simple systems the MSM is the method of choice and has been
highly successful in calculating virial coefficients to quite high
order with impressive precision.

However, when applied to complex flexible molecules such as polymers,
the complexity of implementing computational strategies based on the
Mayer function route seems to increase rapidly
\cite{Caracciolo:2006qf}. For example, to date, calculations of virial
coefficients for molecules have been limited to simple united atom
representations \cite{Shaul:2011fk}, while for flexible molecules such
as polymers, studies have not gone beyond simple lattice walks
\cite{Caracciolo:2006qf,Caracciolo:2008qr,Shida:2007zr,Randisi:2013ly}
or fused hard sphere chains in the good solvent regime
\cite{Vega:2000jk}. The difficulties seem to arise in part from the
need to calculate additional terms in the cluster integral arising
from the molecular flexibility \cite{Shaul:2011fk}, as well as the
lack of an obvious reference system for implementing the MSM in
systems of `soft' particles such as polymers. Accordingly there is
a scarcity of simulation measurements of measurements of third virial
coefficients for complex molecules.

By construction, the Mayer function route requires knowledge of the
interparticle potential. However, in the context of coarse-grained
models there are instances where this potential is unknown but one
nevertheless seeks to measure virial coefficients. Examples are
colloid-polymer mixtures or molecules in explicit solvent which are
often modelled as a highly size-asymmetrical binary mixture. To make
theoretical progress with such systems, one typically considers an
effective one component fluid of the large particles with interactions
determined implicitly by the small ones in which they are
immersed. These effective interactions are inherently many-body in
form, and calculating their form is a tall order. Often, though, it is
possible to perform simulations of the small numbers of the large
particles in a sea of the small ones. The ability to deduce the virial
coefficients of the effective fluid from these simulation would be
useful as it would allow one to predict its likely phase behaviour at
higher colloid density. However, in the absence of the effective
potential, the Mayer function approach cannot do this directly.

\section{Method}
\label{sec:method}

In view of these issues it is interesting to explore alternatives to
the Mayer function route. One such alternative is well known. It is
based on the derivation of the virial expansion from the grand
canonical partition function \cite{Hill:1988ye}; we shall refer to it
as the partition function route. As discussed below, this approach
expresses the $N$th virial coefficient, $B_N$, in terms of sums and
differences of products of the partition function integrals $Z_1\ldots
Z_N$. Heretofore the partition function route has been largely
discounted on dual grounds.  Firstly, to calculate a given virial
coefficient entails the calculation of multiple integrals -- in
contrast to the Mayer formulation which requires only a single
integral. Secondly, it potentially suffers from rapidly deteriorating
precision as $N$ increases because it estimates the relatively small
value of $B_N$ as a difference of large numbers
\cite{Hellmann:2011oq}.

In this section we show that the need to perform explicit numerical
integration of the partition functions $Z_1\ldots Z_N$ can be
circumvented by means of a simple simulation sampling procedure which
directly determines relevant ratios of these integrals via
measurements of the volume-dependent asymptotic value, $f(V)$, of a
{\em structural} function $g_N^\prime(r_{\rm min})$. This function
depends solely on relative molecular position, and as such is as easy
to measure for complex molecules as it is for point particles. By the
same token, it is obtainable by any simulation scheme capable of
generating equilibrium configurations, such as Molecular Dynamics,
Monte Carlo or Langevin Dynamics. Furthermore, since virial
coefficients are derived from structural information rather than
integration of Boltzmann factors, no knowledge of the effective interaction
potential is required. This permits deployment of the method to
subsets of particles, as is relevant for coarse-grained models such as
effective fluids, as discussed above. 

\subsection{Low order virial coefficients of molecular systems}
\label{sec:molecules}

Consider a simulation box of volume $V$ containing $N$ interacting
molecules in thermal equilibrium at inverse temperature
$\beta=(k_BT)^{-1}$. For each of the $N$ molecules we tag an arbitrary
atomic site and label its position vector $\br_i$, with $i=1\ldots N$.
The position vectors of the remaining $m$ atoms in each molecule we
write as $\br_{i,j}=\br_i+\bu_{i,j}, j=1\ldots m$, with $\bu_{i,j}$
the displacement of atom $j$ on molecule $i$ from the tagged atom
$\br_i$. Accordingly a molecular configuration can be specified via a
list of the $N$ tagged and the $M=Nm$ non-tagged coordinates,
$\br^N,\bu^M$. The corresponding Boltzmann probability is
\be
P_N(\br^N,\bu^M)=\frac{ e^{-\beta U(\br^N,\bu^M)}} {Z_N} \:,
\label{eq:boltzmann}
\ee
where $U(\br^N,\bu^M)$ is the full interaction potential containing both intra and intermolecular terms and
\be
Z_N=\int e^{-\beta U(\br^N,\bu^M)}\mathrm{d}\br^N\mathrm{d}\bu^M
\ee
is the $N$-molecule configurational integral.

Now define 
\bea
\tilde{g}_N(\br^N,\bu^M) &\equiv& \frac{P_N(\br^N,\bu^M)}{P_N^{\rm ig}(\br^N)}\\
\:&=& V^N \frac{ e^{-\beta U(\br^N,\bu^M)}}{Z_N}\:,
\label{eq:main}
\eea
where $P_N^{\rm ig}(\br^N)=V^{-N}$ is the probability of finding (within the same volume) a set of $N$ structureless ideal gas particles
in the same configuration as the tagged sites. We shall focus on the low density limit of
$\tilde{g}_N(\br^N,\bu^N)$, corresponding to $|\br_k-\br_l|\to\infty,
\:\;\forall\: k,l$. In this regime the molecules are non-interacting, so we
can integrate out the internal molecular degrees of freedom
(associated with the $\bu_{i,j}$) to obtain the asymptotic value
\be
f_N(V)\equiv\lim_{|\br_k-\br_l|\to\infty}\tilde{g}_N(\br^N)=\frac{(\Omega V)^N}{Z_N}=\frac{Z_1^N}{Z_N}\:,
\label{eq:limit}
\ee
where $\Omega$ is the integral over the internal degrees of freedom of
a single molecule (which is equal to $8\pi^2$ for a rigid molecule with no special symmetries) and
$Z_1$ is the corresponding configurational integral.

The quantity $f_N(V)=Z_1^N/Z_N$ is central because it permits a direct
calculation of molecular virial coefficients as will be shown below. A
key feature is its dependence on the system volume. Specifically,
although it has the limiting behaviour $\lim_{V\to\infty} f_N(V)=1$,
(because $Z_N$ is dominated by configurations in which the molecules are well separated)
for finite system volume $f_N(V)$ deviates from unity.  Clearly, however, determining $f_N(V)$ by simulation via eq.~\ref{eq:limit}
is not a feasible proposition since it entails populating a
$3N$-dimensional histogram for $P_N(\br^N)$ with sufficient
statistics to yield precise probabilities. Fortunately, though, it
turns out to be possible to determine $f_N(V)$ using only
one-dimensional histograms. To see this, consider the quantity
\be
g^\prime_N(r_{\rm min})\equiv\frac{P_N(r_{\rm min})}{P_N^{\rm ig}(r_{\rm min})}\:.
\label{eq:gNrmin}
\ee
Here $r_{\rm min}$ is, for some configuration, the magnitude of that
vector between a pair of the $N$ tagged sites that is smaller than all
other separation vectors. In the course of a simulation, one can
accumulate histograms for $P_N(r_{\rm min})$ and $P_N^{\rm ig}(r_{\rm
  min})$ and thus form $g^\prime_N(r_{\rm min})$. We note that
$P_N^{\rm ig}(r_{\rm min})$ is particularly simple to measure because
it simple involves repeatedly picking $N$ random points within the box
volume $V$ and finding the shortest of the $N(N-1)/2$ pair separations.

Now clearly the limit $r_{\rm min}\to\infty$ is none other than the
limit $|\br_k-\br_l|\to\infty, \:\;\forall\; k,l$. Moreover, since in
this limit the microstates of the tagged particles are visited with
equal probability $\Omega^N Z_N^{-1}$, while those of the ideal gas are
visited with equal probability $V^{-N}$, it follows that the
limiting value of $g^\prime_N(r_{\rm min})$ is the same as that of
$\tilde{g}_N(\br^N)$, i.e.
\be
\lim_{r_{\rm min}\to \infty}g^\prime_N(r_{\rm min})= f_N(V)\:.
\label{eq:gNrminlim}
\ee

Equation~(\ref{eq:gNrminlim}) provides a straightforward computational prescription for
determining $f_N(V)$, which in turn permits the calculation of the
virial coefficients for the molecular system. For instance from the virial cluster expansion (see Appendix) one finds that for $N=2$
particles

\bea
B_2 &=& \frac{V}{2}(1-\frac{Z_2}{Z_1^2})\nonumber\\
    &=& \frac{V}{2}\left( 1-\frac{1}{f_2(V)} \right )\:,
\label{eq:B2}
\eea
Similarly for three particles one has 
\bea
B_3 &=& \frac{V^2(Z_1^4-3Z_2Z_1^2-Z_3Z_1+3Z_2^2)}{3Z_1^4}\nonumber\\
%    &=& \frac{ V^2}{3} \left(-\frac{1}{f_3}-\frac{3}{f_2}+\frac{3}{f_2^2}+1\right)\\
    &=& 4B_2^2-2B_2V+V^2\frac{[f_3(V)-1]}{3f_3(V)}\:.
\label{eq:B3}
\eea

While for four particles one finds 
\begin{multline}
B_4 = \frac{V^3}{8 Z_1^6}\left(2 Z_1^6-12 Z_2 Z_1^4-8 Z_3 Z_1^3+\left(27 Z_2^2-Z_4\right) Z_1^2\right.\\
              \left.+12 Z_2 Z_3 Z_1-20 Z_2^3\right)\\
%B_4 = \frac{V^3}{8} \left(-\frac{8}{f_3}-\frac{1}{f_4}+\frac{12}{f_3 f_2}-\frac{12}{f_2}+\frac{27}{f_2^2}-\frac{20}{f_2^3}+2\right)\\
B_4 =  \frac{1}{8} \left(-12 B_2 \left(V^2-6 B_3\right)+60 B_2^2 V-12 B_3 V\right.\\
\left.-128 B_2^3+\left[1-1/f_4(V)\right] V^3\right)\:.
\label{eq:B4}
\end{multline}
More generally, knowledge of $Z_i, i=2,\ldots, m$ permits the calculation of the $m$th virial coefficient $B_m$.

\subsection{Isolating three-body interactions in coarse-grained effective models}

\label{sec:B3compare}

The ability to measure second and third order virial coefficients in
molecular systems provides a route to quantifying the role of
three-body interactions in effective coarse-grained models. To appreciate
this, consider the third virial coefficient $B_3^{\rm mol}$ of the
fully detailed molecular system. This contains information on both two
and three-body interactions that appear in the full effective
potential \cite{Hansen-MacDonald}. In order to isolate the three-body
contribution one can compare $B_3^{\rm mol}$ for the full molecular
system with $B_3^{\rm pair}$, the third virial of a system of $N=3$
particles interacting via an effective pair potential (potential of
mean force) \cite{DAdamo:2012ys}. Now if, by construction, both the
full molecular model and the effective pair potential have the same
value of $B_2$, then the difference $B_3^{\rm mol}-B_3^{\rm pair}$
clearly isolates the contribution of (non-additive) three-body interactions to
$B_3^{\rm mol}$.

Such a comparison is effected very naturally within our method because
the function $g_2^\prime(\rmin)$ (which is just $g_2^\prime(\rmin)$
for two particles) not only provides an estimate of $B_2$, it also
yields the effective pair potential:

\be
\beta W^{\rm pair}(r)=-\ln[g_2^\prime(r)/f_2(V)]\:.
\ee
Thus measurements of $g_2^\prime(r)$ and thence $f_2(V)$ for some $V$ can be used to
determine $W^{\rm pair}(r)$  \cite{Ashton:2011kx}. A simulation of three particles
interacting via this pair potential provides an estimate of $B_3^{\rm
  pair}$, which can be compared with that arising from a simulation of
three molecules.

\subsection{Extension to colloid-polymer mixtures}
\label{sec:collpol}

The formalism for molecular systems can be readily adapted to mixtures
in which the coarse-graining involves tracing out the degrees of
freedom associated with one species.  Common examples are molecules in
solution and colloid-polymer mixtures. For definiteness we shall
specialize to the latter case which, as noted above is often
modelled as a highly size-asymmetrical mixture of spheres. The
coarse-graining procedure integrates out the small particle degrees of
freedom to yield an effective one component fluid of the large
spheres.  The statistical mechanics of these spheres is exactly
described via an effective Hamiltonian \cite{Dijkstra1999}

\be
H^{\rm  eff}=H^{0}+\Theta\:.
\ee
Here $H^0$ is the bare interaction between the large particles
while $\Theta$ is a many-body contribution arising from the small
particles which can in turn be written as a sum over $n$-body terms

\be
\Theta=\sum_{n=1}^\infty\theta_n\:.
\label{eq:theta}
\ee
Such many-body terms arise from the coarse-graining procedure even when the underlying interactions are pairwise additive.

Now for a system of $N$ large particles described by this effective
Hamiltonian, the configurational statistics are given by

\be
P_N(\br^N)=\frac{ e^{-\beta H^{\rm eff}(\br^N)}} {Z_N} \:,
\ee
where 
\be
Z_N=\int e^{-\beta H^{\rm eff}(\br^N)}\dbr^N
\ee
is the partition function of the effective fluid of $N$-colloids.
So by analogy with the arguments of Sec.~\ref{sec:molecules}, when the large particles are all well separated one has
\be
\lim_{r_{\rm min}\to\infty}{g}^\prime_N(\br^N)=f_N(V)=\frac{V^N}{Z_N}\:,
\label{eq:mainmixture}
\ee
because $Z_1=V$.

Generally speaking, the full many-body effective Hamiltonian is
inaccessible due to the analytical and computational difficulties of
calculating its form exactly.  In its absence, it is common practice
in theoretical treatments to resort to a pair potential approximation,
ie. to truncate the series for $\Theta$ (Eq.~\ref{eq:theta}) at
$n=2$. For colloid-polymer mixtures the resulting pair potential is
known as the depletion potential. To assess the neglect of three-body
interactions in this approximation on the thermodynamics of the system,
one can follow the strategy of Sec.~\ref{sec:B3compare} and measure
the difference $B^{\rm eff}_3-B^{\rm dep}_3$. To do so one first
apples the method of Sec.~\ref{sec:molecules} in simulations of a
system of $N=3$ particles interacting via the depletion pair potential
to yield $B^{\rm dep}_3$. Next one simulates $N=3$ large particles
immersed in the sea of small ones (a task for which specialized
algorithms may be required, see Sec.~\ref{sec:mixture}). But by {\it
  fiat}, the statistical mechanics of the large particles in such a
simulation are just those of the effective system. Accordingly if in
such a simulation one tags the large particles, then
Eq.~\ref{eq:mainmixture} together with Eqs.~\ref{eq:B2} and \ref{eq:B3}
provide estimates of $B^{\rm eff}_3$.

\section{Commentary}
\label{sec:commentary}
\subsection{Optimization and limitations}
\label{sec:optimise}

Our method for estimating virial coefficients rests on measurements of
the asymptote $f_N(V)$. Generally speaking, for a given computational
expenditure, the numerical precision of the resulting estimates for
$B_N$ can be optimised by choosing as small a system volume $V$ as
possible consistent with maintaining access to the asymptotic regime of
$g^\prime_N(r_{\rm min})$. To appreciate this, consider the case of
the absolute error in $B_2$. From Eq.~(\ref{eq:B2}) this is

\be
\delta B_2=\frac{V}{2f_2^2}\delta f_2\:,
\ee
showing that an absolute error $\delta f_2$ in $f_2$ is scaled up by a factor $V$. As far as the relative error is concerned one has
\be
\frac{\delta B_2}{B_2}=\frac{\delta f_2}{f_2(f_2-1)}\:,
\ee
which shows that this is sensitive to the magnitude of the finite-size `signal' $f_2(V)-1$. However, from Eq.~(\ref{eq:B2}),
\be
f_2(V)-1=\frac{2B_2}{V-2B_2}\:,
\ee
and since $B_2$ is fixed by the model, this shows that in order to obtain a larger signal, it helps to choose a small $V$.

Similar arguments relate to $B_N$ with $N>2$,  though here the absolute error
grows like $V^{N-1}$, which implies that a large computational
investment is required to access virial coefficients higher than the
third. One is helped, however, if the interactions are short ranged
since this allows access to the asymptotic regime of
$g^\prime_N(r_{\rm min})$ using small system volumes.  We also note that for
systems with attractive interactions, the magnitude of $f_N(V)-1$ is
increased by lowering the temperature so that the
molecules spend more time in close contact.  Although at very low
temperatures this effect could potentially result in poor statistics
for estimates of $f_N(V)$, this problem can be easily surmounted by
using biased (umbrella) sampling to enhance the sampling in the tail
region of $P_N(r_{\rm min})$.

\subsection{Radial distribution functions,  tail effects and pair potentials}
\label{sec:rdf}

The radial distribution function $g(r)$ is a key structural
characteristic of a fluid system and  a common goal of simulations is to
measure its form accurately.  However, simulation estimates
are necessarily based on a finite number of particles $N$ in a finite
volume $V$. As is well known, in contrast to the true $g(r)$, the
asymptote of the measured function (let us denote it $g_N(r)$) fails
to approach unity even at infinite volume
\cite{Hansen-MacDonald,Lebowitz:1961uq,Lebowitz:1961fk,Perera:2011fk,Kezic:2012ly,Kolafa:2002uq,Koga:2013kx}. This
`tail effect' complicates the extraction of structural and thermodynamic information
from simulation measurements of $g_N(r)$ which has to be corrected by
an empirical scaling procedure. 

The finite-$N$ dependence of the asymptote of $g_N(r)$ arises from the
fact that the definition of $g(r)$ in terms of the two body
distribution function is normalized using the density of an ideal gas
of $N$ rather than $N-1$ particles \cite{Hansen-MacDonald}. This definition is followed
in algorithmic prescriptions for measuring $g_N(r)$ by simulation
\cite{Frenkelsmit2002,AllenTildesley}. However the considerations of
Sec. ~\ref{sec:molecules} suggests a better finite-size estimator
for $g(r)$, namely:

\begin{equation}
g^\prime(r)\equiv\frac{P(r)}{P^{ig}(r)}\:,
\end{equation}
where $P(r)$ is the probability of finding a particle a
distance $r$ from another particle (assumed to be at the origin) in the $N$ particle system and $P^{ig}(r)=4\pi r^2/V$. 

$g^\prime(r)$ is simply related to the standard definition by 

\be
g^\prime(r)=\frac{N}{N-1}g_N(r)\:.
\ee
But while both $g^\prime(r)$ and $g_N(r)$ will be identical 
in the thermodynamic limit, $g^\prime(r)$ provides a
superior finite-size estimator because it obviates the
need to deal with corrections to the asymptote associated with finite $N$. Specifically
its asymptote is unity in the limit $V\to\infty$ for all $N>1$. Accordingly, in simulations one only
has to correct $g^\prime(r)$ for finite-{\em volume} effects, and the associated degree of empirical
scaling will thus be generally less than for $g_N(r)$.

The differences between the two estimators will of course be small in
most situations because generally one deals with hundreds or thousands
of particles. However when seeking to calculate pair potentials (which
are defined in the limit of vanishing density), one considers only a
pair of particles, $N=2$. For this case, $g^\prime(r)=2g_2(r)$ -- a
stark difference. Since for a pair of particles $r_{min}=r$,
this is just the situation considered in detail in
Sec~\ref{sec:molecules} which sets out the relationship between the
finite-$V$ tail effect and the second virial coefficient.

\subsection{Relationship to the Kirkwood-Buff integrals for mixtures}

Although a slight digression, it is notable that a variation of our method 
for measuring the second virial coefficient provides a route to 
estimating the Kirkwood-Buff integrals which quantify the net affinity of a pair of
species in a fluid mixture. For components $i$ and $j$ these integrals are
defined

\be
G_{ij}=\int_0^\infty (g_{ij}(r)-1)4\pi r^2 \mathrm{d}r\:,
\label{eq:kbi}
\ee
where $g_{ij}(r)$ is the partial radial distribution function. 

Although for large numbers of particles the effects of finite $N$ on
$g_{ij}(r)$ are much less pronounced than for the case of
$N=2$ discussed above, the asymptote of $g_{ij}(r)$ will nevertheless
typically differ significantly from unity in simulations. In practice
this complicates accurate measurements of Kirkwood-Buff integrals
\cite{Perera:2011fk,Kruger:2013wq,Matteoli:1995kx}.

It is therefore noteworthy that estimates of the
Kirkwood-Buff integrals can be obtained without the need for
integration by measuring the asymptotic value

\be
f_{ij}(V)\equiv\lim_{r_{ij}\to\infty}g^\prime_{ij}(r_{ij})
\ee
where 
\be
g^\prime_{ij}(r_{ij})=\frac{P(r_{ij})}{P^{ig}(r_{ij})}\:,
\label{eq:kbigprime}
\ee
with $P(r_{ij})$ the probability of finding a particle of species $j$ a
distance $r_{ij}$ from a particle of
species $i$ (the latter assumed to be at the origin) and $P^{ig}(r_{ij})=4\pi r^2/V$. 

If one inserts Eq.~\ref{eq:kbigprime} into Eq.~\ref{eq:kbi} instead of
$g_{ij}(r_{ij})$ one finds $G_{ij}=0$ always. The reason for this is a
finite-size effect, namely the fact that the asymptote $f_{ij}(V)$ of
$g^\prime_{ij}(r_{ij})$ differs from unity. In order to obtain the
correct integral, one therefore has to multiply $g^\prime_{ij}$ by a
factor $f_{ij}{(V)}^{-1}$. It thus follows
that the Kirkwood-Buff integral can be written in terms of the measured asymptote $f_{ij}(V)$, ie.

\be
G_{ij} = V\left(\frac{1}{f_{ij}(V)}-1 \right )\:,
\label{eq:ourKBI}
\ee
which is analogous to Eq.~\ref{eq:B2}.

\section{Illustration and test}
\label{sec:tests}

\begin{figure}[h]
\includegraphics[type=pdf,ext=.pdf,read=.pdf,width=0.9\columnwidth,clip=true]{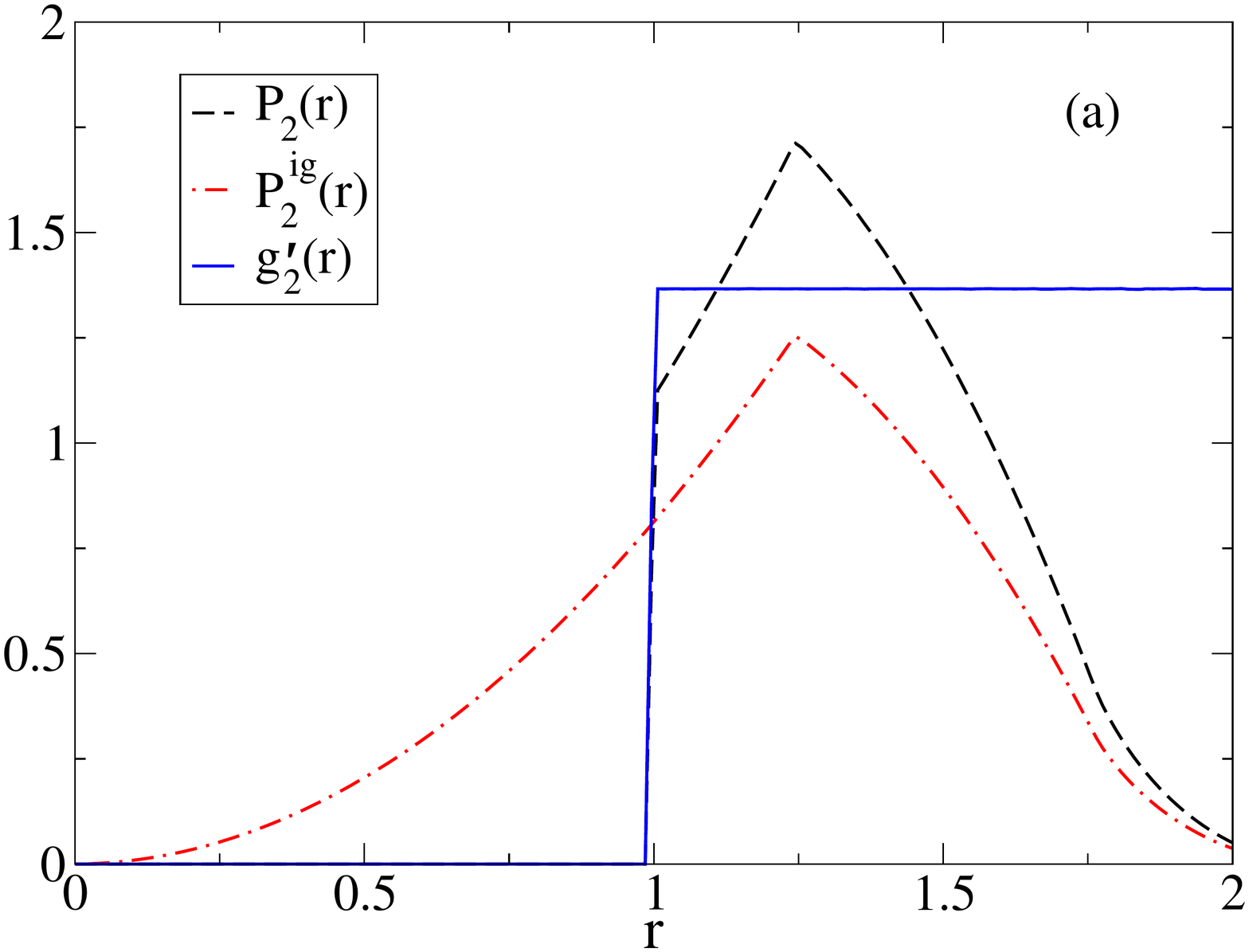}
\includegraphics[type=pdf,ext=.pdf,read=.pdf,width=0.9\columnwidth,clip=true]{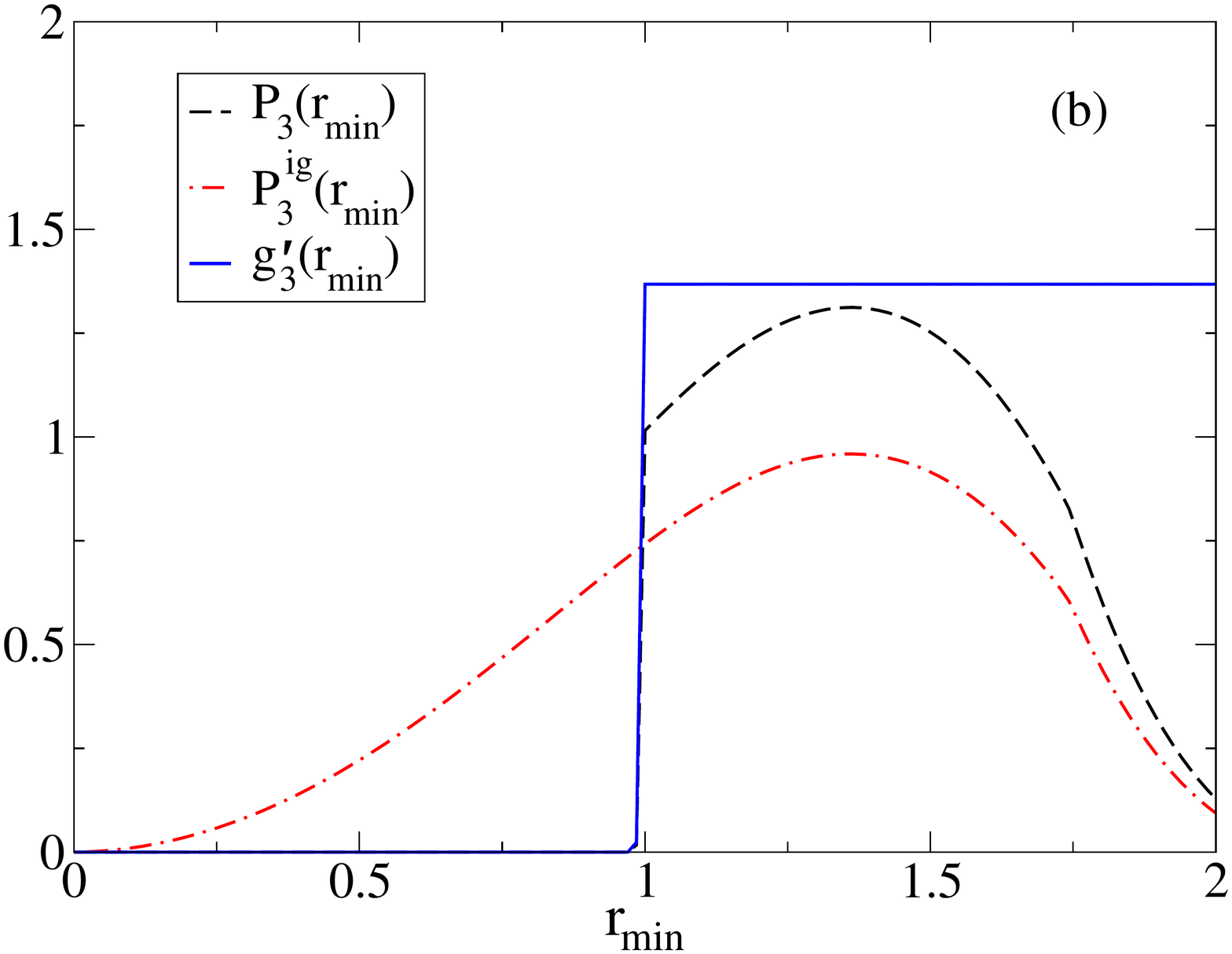}
 \caption{{\bf (a)} Simulation estimates of $P_2(r)$, $P_2^{ig}(r)$
   and their ratio $g^\prime_2(r)$ as obtained in an MC
   simulation with a periodic box of volume $V=(2.5\sigma)^3$. {\bf (b)} Estimates of $P_3(r_{\rm min})$,
   $P_3^{ig}(r_{\rm min})$ and their ratio $g^\prime_3(r_{\rm min})$ obtained
   in MC simulations with a periodic box of size $V=(3.5\sigma)^3$. }
 \label{fig:test}
 \end{figure}

In order to both illustrate and test the method, it is instructive to
use it to estimate the first few virial coefficients of a single
component system of hard spheres, for which exact values of virial
coefficients are known. Considering first the calculation of $B_2$ for
hard spheres, here one has $N=2$ and hence $r_{\rm min}=r$.
Fig.~\ref{fig:test}(a) plots the measured forms of $P_2(r)$,
$P_2^{ig}(r)$ and their ratio $g^\prime_2(r)$ obtained by a simple
Monte Carlo simulation of a pair of particles in a periodic box
of volume $V=(2.5\sigma)^3$, with $\sigma$ the hard sphere
diameter. For this simple interaction potential, the limiting value of
$g^\prime_2(r)$ pertains for all $r>\sigma$. Note also that both the
probability distributions $P_2(r)$ and $P_2^{ig}(r)$ show a maximum as
$r_{\rm min}$ increases. This is a finite-size effect arising from the
fact that the available volume at large $r_{\rm min}$ decreases due to
the cubic box geometry. Nevertheless $g^\prime_2(r)$ remains flat
inside this regime because the contribution from
this finite-size effect exactly cancels when forming the ratio of the two
distributions.

In general one can estimate $f_N(V)$ visually, or from a fit. However,
we have found that a particularly accurate measure results from the
ratio of integrals

\be
f_N(V)=\frac{\int_{r_l}^{r_u} P_N(r_{\rm min})\mathrm{d}r_{\rm min}}{\int_{r_l}^{r_u} P^{\rm ig}_N(r_{\rm min})\mathrm{d}r_{\rm min}}\:,
\label{eq:findf}
\ee
where $r_l$ is some value of $r_{\rm min}$ for which $g^\prime(r_{\rm
  min})$ can be considered to have first reached its limiting value, and
$r_u$ is the largest value of $r_{\rm min}$ for which data has been
accumulated, which will typically be half the simulation box diagonal length. It should be emphasized that in practice,
eq.~(\ref{eq:findf}) is evaluated simply from a count of entries in the
respective histograms for $P_N(r_{\rm min})$ and $P_N^{\rm ig}(r_{\rm
  min})$--no numerical quadrature is performed.  In this way we find (taking $r_l=\sigma$) that 
$B_2=2.09441(6)\sigma^3$, to be compared with the exact value
$B_2=2\pi\sigma^3/3=2.094395\sigma^3$.

Turning next to the third virial coefficient $B_3$, here one has $N=3$
hard spheres and MC simulations of a trio of particles in a box of
volume $V=(3.5\sigma)^3$ yields the forms of $P_3(r_{\rm min})$,
$P_3^{ig}(r_{\rm min})$ and $g^\prime_3(r_{\rm min})$ shown in
Fig.~\ref{fig:test}(b). Again for this potential, $g^\prime_3(r_{\rm
  min})$ reaches its limiting value immediately for $r_{\rm
  min}>\sigma$ and we find via eq.~(\ref{eq:B3}), that
$B_3=2.7418(4)\sigma^6$. This is to be compared with the exact value of
$B_3=5\pi^2\sigma^6/18=2.741557\sigma^6$.

Finally, we consider $B_4$, which we have measured for a system of size $V=(3.5\sigma)^3$.
The measured value of $f_4$ together with Eq.~\ref{eq:B4} implies that $B_4  = 2.629(22)\sigma^9$ to be
compared with the exact value \cite{Clisby:2004lh} of $2.6362\ldots\sigma^9$.

\section{Application I: Three-body interactions in highly size-asymmetrical fluid mixtures}

\label{sec:mixture}

Having validated our method, we now turn to a more challenging
problem, namely that of determining the second and third virial
coefficients of the effective Hamiltonian for highly size asymmetrical
mixtures.

When colloids are immersed in a sea of small particles such as
polymers or much smaller colloids, the small particles mediate
effective colloidal interactions. In the simplest case in which all
particles are hard, the effective interaction arises solely from
entropic effects and results in an enhanced attraction between the
colloids known as the depletion interaction which acts on a length
scale set by the small particle diameter. More generally the nature of
the effective interaction can depend sensitively on the detailed form
of the interactions between the small particles and the large ones.

As described in Secs.~\ref{sec:intro} and \ref{sec:collpol},
theoretically one would like to describe such a system in terms of a
single component model of colloids interacting according to  a many-body effective
Hamiltonian. This Hamiltonian would in principle provide an exact representation
of the actual mixture and contain two-body, three-body, etc terms. However, many-body
terms can be difficult to calculate in practice and thus one usually
focuses on the two-body contribution in the expectation that (at least
for small $q$), many-body effects should be small.

A commonly studied model treats the colloids as big hard spheres of
diameter $\sigma_b$ and the small particles as hard spheres of
diameter $\sigma_s$. The size ratio is then expressed as
$q\equiv\sigma_s/\sigma_b$. In the two-body approximation, the
effective Hamiltonian between the colloids is written as a sum over
pair interactions:
\be
 H^{\rm  eff}\approx\sum_{i,j}[\phi(r_{ij})+W(r_{ij})]\:,
\ee
where $\phi(r_{ij})$ is the hard sphere interaction between a pair of colloids
whose centers are separated by a distance $r_{ij}$, while $W(r_{ij})$ is the
depletion pair potential, whose form depends on the small particle volume fraction and
model details such as whether or not the big-small interaction is additive.
Typically one imagines that the small particles are in
equilibrium with a reservoir so that $W(r)$ is parameterized in terms
of the {\em reservoir} volume fraction $\eta_s^r$.

Depletion potentials have been estimated using both theory and
simulation.
\cite{Gotzelmann1998,Roth2000a,Roth2001,Louis2001,Louis2002a,Louis:2002fk,Roth:2001vn,Yuste2008,Amokrane2005,Ayadim2006,Oettel2004,Oettel2009,Botan2009,Ashton:2011kx,Ashton:2013kq}
for a variety of models such as additive and non-additive hard
spheres. Calculations of the phase behaviour of systems interacting
through depletion potentials has revealed interesting features such as a
metastable fluid-fluid phase separation at high values of $\eta_s^r$
\cite{Dijkstra1999,Rotenberg2004,Largo2006} in additive hard spheres. However since
simulations of depletion potentials neglect the many-body forces that
occur in the full mixture, the question as to whether such effects
actually occur in the full mixture is somewhat moot
\cite{Barrat1986,Haro:2013bh}.

In view of this it is desirable to have techniques for quantifying the
effects of neglecting many-body interactions when employing depletion
potentials. Let us focus on three-body interactions, which are
(typically) the dominant many-body effect. As discussed in
Sec.~\ref{sec:B3compare}, one systematic way to study three-body
effects in this system (there are others
\cite{Goulding:2001fk,Malherbe2001,Amokrane2003}) is to calculate the
third virial coefficient $B_3^{\rm eff}$ for the full effective fluid
and compare it with the corresponding value $B_3^{\rm dep}$ arising
from a simulation of three particles interacting via the appropriate
depletion potential \footnote{Note that neither of these quantities is
  to be confused with the third virial coefficient of the full mixture
  \protect\cite{Enciso:1998ys}, which is a quantity which treats small
  and large particles on an equal footing.}. This comparison will
directly probe the extent to which the net interaction between a pair
of large hard spheres particles in the full mixture is influenced by
the proximity of a third large sphere which modulates the small
particle density distribution. By contrast for a trio of particles
interacting through the two-body depletion potential, no such effect
exists by definition.

Below we describe how one can perform this comparison by combining the
method described in Sec.~\ref{sec:method} for estimating virial
coefficients, with state-of-the-art simulation techniques for dealing
with highly size asymmetrical fluid mixtures.

\subsection{Computational procedure for determining $B_3^{\rm eff}$ and $B_3^{\rm
  dep}$}
\label{sec:procedure}

In order to obtain estimates for both $B_3^{\rm eff}$ and $B_3^{\rm
  dep}$ in highly size asymmetrical mixtures, we deploy the
geometrical cluster algorithm (GCA) \cite{Dress1995,Liu2004}. This is
a collective updating Monte Carlo scheme which provides efficient
relaxation at practically any particle size ratio provided the overall
volume fraction is not too high. Our implementation reproduces
conditions considered in many experimental and theoretical studies of
colloids, in that we treat the small particles grand canonically so
that their properties are parameterized in terms of the reservoir
volume fractions $\eta_s^r$. Transfers of small particles are effected
using a standard grand canonical approach \cite{Frenkelsmit2002}.
However to utilize this ensemble one needs to know accurately the
chemical potential corresponding to a given $\eta_{\rm s}^r$. For some
types of small particle fluid, such as hard spheres, this relationship
is independently known. Otherwise, it has to be determined in a
separate simulation.

Using the GCA, we can study the statistical mechanics of either a pair
or a triple of big particles in a sea of small ones. The procedure for
implementing the strategy of Sec.~\ref{sec:B3compare} is as follows

\begin{enumerate}

\item[(i)] From simulations of $N=2$ big particles, measure the
  form of $g^\prime_2(r)$ at some prescribed $\eta_s^r$. This yields
  the value of $B_2^{\rm eff}(\eta_s^r)$ via eq.~(\ref{eq:B2}).

\item[(ii)] Use the form of $g^\prime_2(r)$ obtained in (i) 
  to estimate the depletion potential as 
\[
\beta W(r|\eta_s^r) =-\ln[g_2^\prime(r)/f_2(V)]
\]

\item[(iii)] Next simulate three big particles at the same value of
  $\eta_s^r$ and measure $g^\prime_3(r_{\rm min})$. Together
  with the estimate of $B_2^{\rm eff}(\eta_s^r)$ obtained in (i), this provides 
  an estimate for $B_3^{\rm eff}(\eta_s^r)$ via eq.~(\ref{eq:B3}).

\item[(iv)] Finally perform a simple MC simulation of three particles
  interacting via the appropriate depletion potential $W(r|\eta_s^r)$
  as obtained in (ii) to determine the form of $\tilde{g}_3(r_{\rm
    min})$.  Together with the estimate of $B_2^{\rm
    eff}(\eta_s^r)=B_2^{\rm dep}(\eta_s^r)$ obtained in (i), this
  gives the third virial coefficient $B_3^{\rm dep}(\eta_s^r)$ 
  via eq.~(\ref{eq:B3}).

\end{enumerate}

In what follows, we detail our measurements of $B_3^{\rm eff}(\eta_s^r)$ and
$B_3^{\rm dep}(\eta_s^r)$ for two models of highly size-asymmetrical
mixtures, namely the Asakura-Oosawa model and a system of additive hard spheres.

\subsection{Asakura-Oosawa model}

The Asakura-Oosawa (AO) model describes colloidal hard-spheres in a solvent of
non-interacting particles modelling ideal polymer that have a
hard-particle interaction with the colloids
\cite{Asakura1954,Asakura:1958uq}. Owing to its extreme
non-additivity, the model is somewhat analytically tractable. 
Specifically,  the exact form of the depletion pair potential is known
\cite{Asakura:1958uq} to be

\begin{widetext}
\be
\beta W_{\rm AO}(r)=\left \{ \begin{array}{ll}
-\eta_s^r\frac{(1+q)^3}{q^3}\left[1-\frac{3r}{2\sigma_b(1+q)}+\frac{r^3}{2\sigma_b^3(1+q)^3}\right]\mbox{\hspace{1mm}},   &  \sigma_b<\
 r< \sigma_b+\sigma_s \\
                               &       \\
 0, & r\ge  \sigma_b+\sigma_s \, ,\\
\end{array}
\right.
\label{eq:AOpot}
\ee
\end{widetext}
where $\sigma_s$ is the `polymer' diameter, i.e. the colloid-polymer
pair potential is infinite for $r<(\sigma_b+\sigma_s)/2$. This fact
obviates the need to implement steps (i) and (ii) in the procedure of
Sec.~\ref{sec:procedure}.  Another interesting aspect of the
model is that owing to the lack of polymer-polymer interactions,
the effective potential contains no many-body interactions for size ratios $q<0.1547$
\cite{Dijkstra:1999kl}. This renders the model an excellent proving ground
for testing the sensitivity of our approach.

\begin{figure}[h]
\includegraphics[type=pdf,ext=.pdf,read=.pdf,width=0.95\columnwidth,clip=true]{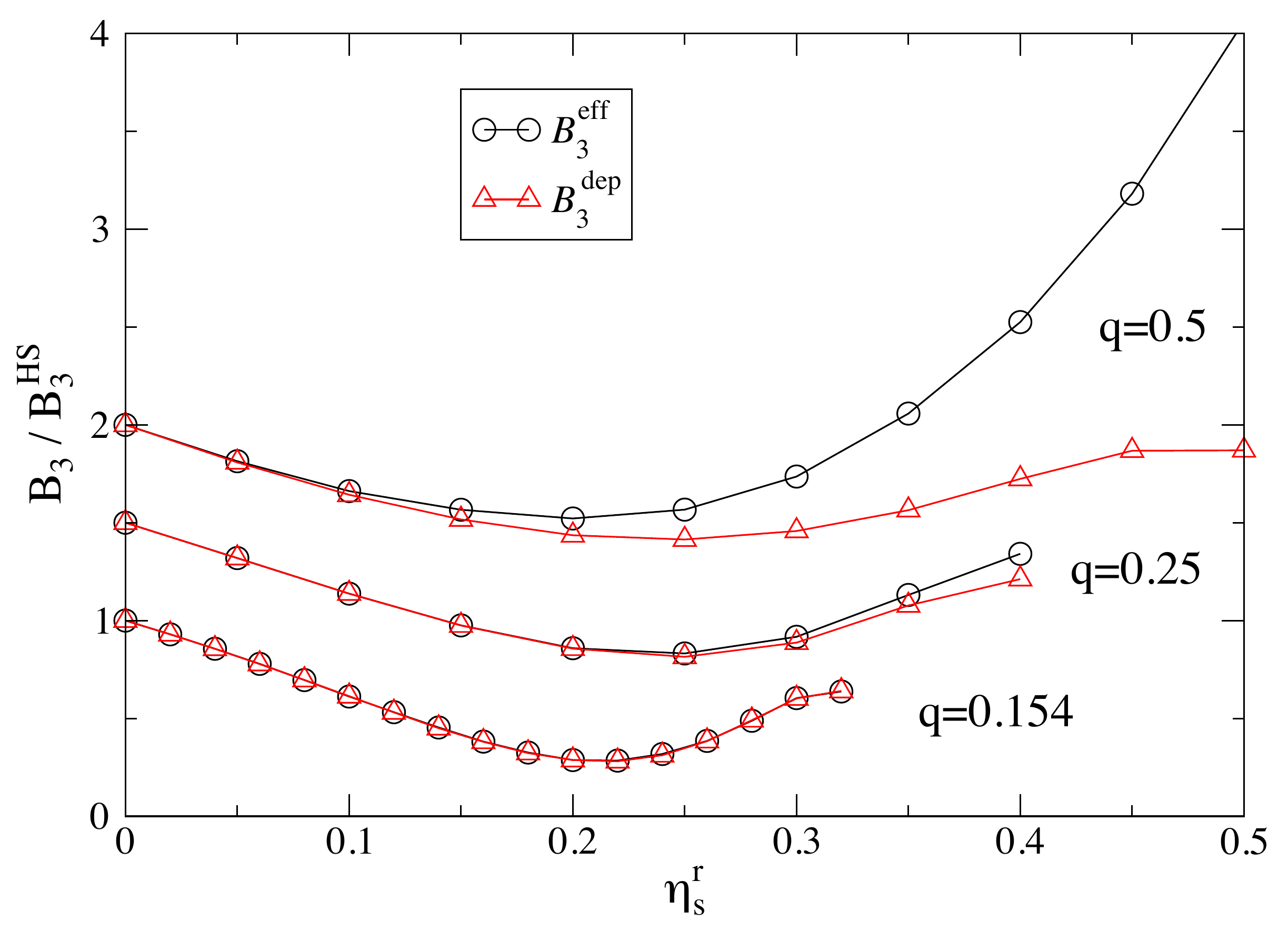}
\caption{Comparison for size ratios $q=0.5,0.25, 0.154$ and various
  $\eta_s^r$ of the measured values of $B_3$ arising from the
  simulation of the full two component AO model ($B_3^{\rm
    eff}(\eta_s^r)$), and a three particle system interacting via the
  AO depletion potential of eq.~(\ref{eq:AOpot}) ($B_3^{\rm
    dep}(\eta_s^r)$). Lines are guides to the eye and statistical uncertainties are smaller than the
  symbol sizes. To aid visibility, the curves for $q=0.25$ and $q=0.5$ have been shifted vertically by $0.5$ and $1.0$ respectively.}

 \label{fig:B3ao}
 \end{figure}

Fig.~\ref{fig:B3ao} shows our estimates of $B_3^{\rm eff}(\eta_s^r)$
and $B_3^{\rm dep}(\eta_s^r)$ for three size ratios, $q=0.5, 0.25,
0.154$ obtained using the methodology of Sec.~\ref{sec:method}.  The
estimates of $B_3$ are all normalized by the value for pure hard
spheres which pertains in the limit $\eta_s^r\to 0$, and curves for
the various $q$ are shifted for clarity. One expects that many
body effects, quantified by the difference between $B_3^{\rm
  eff}(\eta_s^r)$ and $B_3^{\rm dep}(\eta_s^r)$ should increase with
$\eta_s^r$ and this is indeed the case, as our data show. Furthermore,
the differences should diminish with decreasing $q$ and have
disappeared by $q=0.154$. Again this is confirmed by our data.

\subsection{Additive hard sphere mixtures}

Turning now to the more challenging system of additive hard spheres,
here the GCA permits access to a much more limited range of $\eta_s^r$
than for the AO model, and beyond $\eta_s^r=0.2$ relaxation times
become prohibitive. In contrast to the AO model the depletion pair
potential is not known exactly for the additive mixtures, so we
measure it in a simulation of two large particles as described in
Sec.~\ref{sec:procedure}.  For this system the reservoir chemical
potential $\mu(\eta_s^r$) of the small particles is obtained from the
equation of state of Kolafa {\em et al}\; \cite{Kolafa:2004vn}, which
we have checked provides a highly accurate representation of grand
canonical ensemble simulation data.

\begin{figure}[h]
\includegraphics[type=pdf,ext=.pdf,read=.pdf,width=0.95\columnwidth,clip=true]{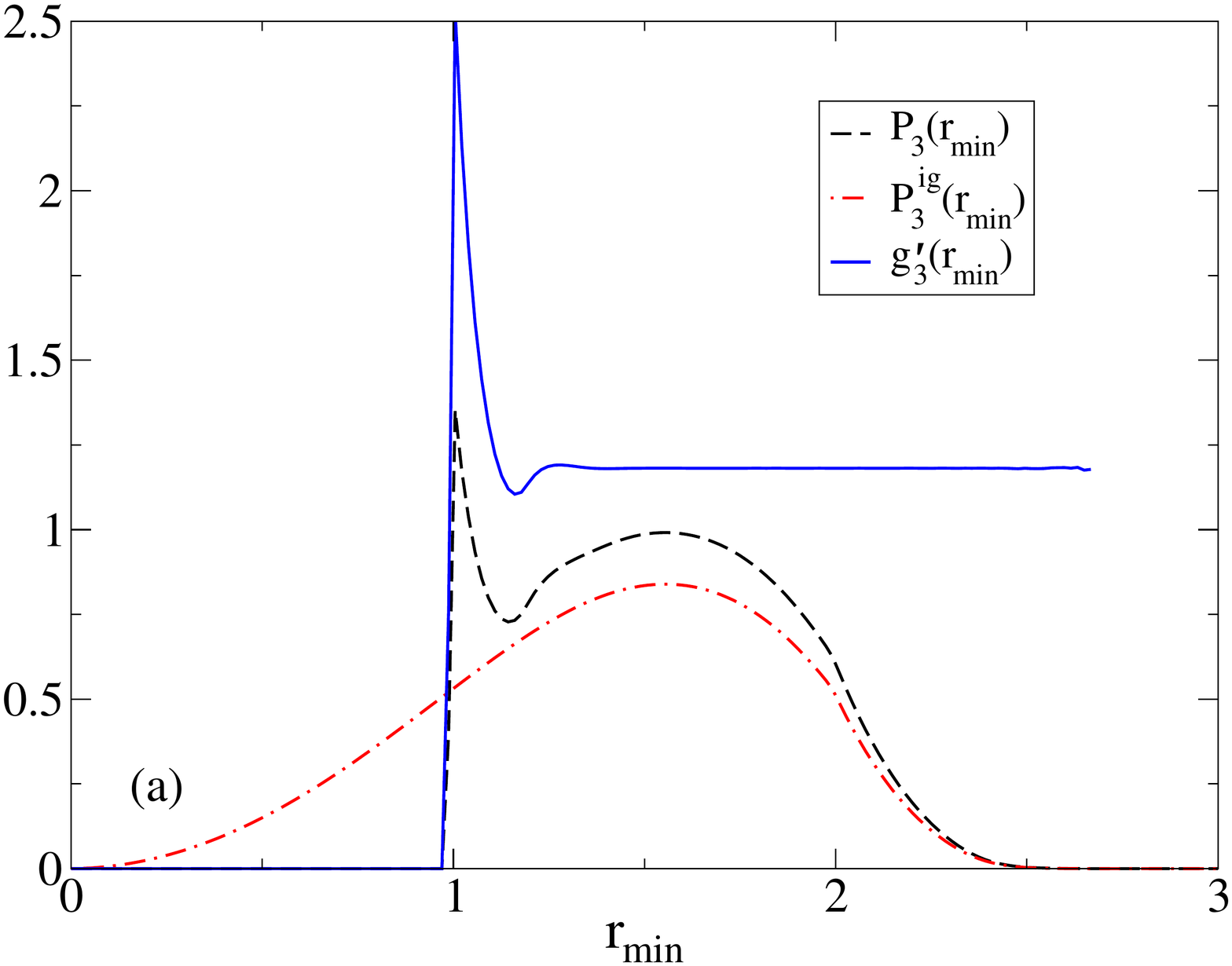}
\includegraphics[type=pdf,ext=.pdf,read=.pdf,width=0.95\columnwidth,clip=true]{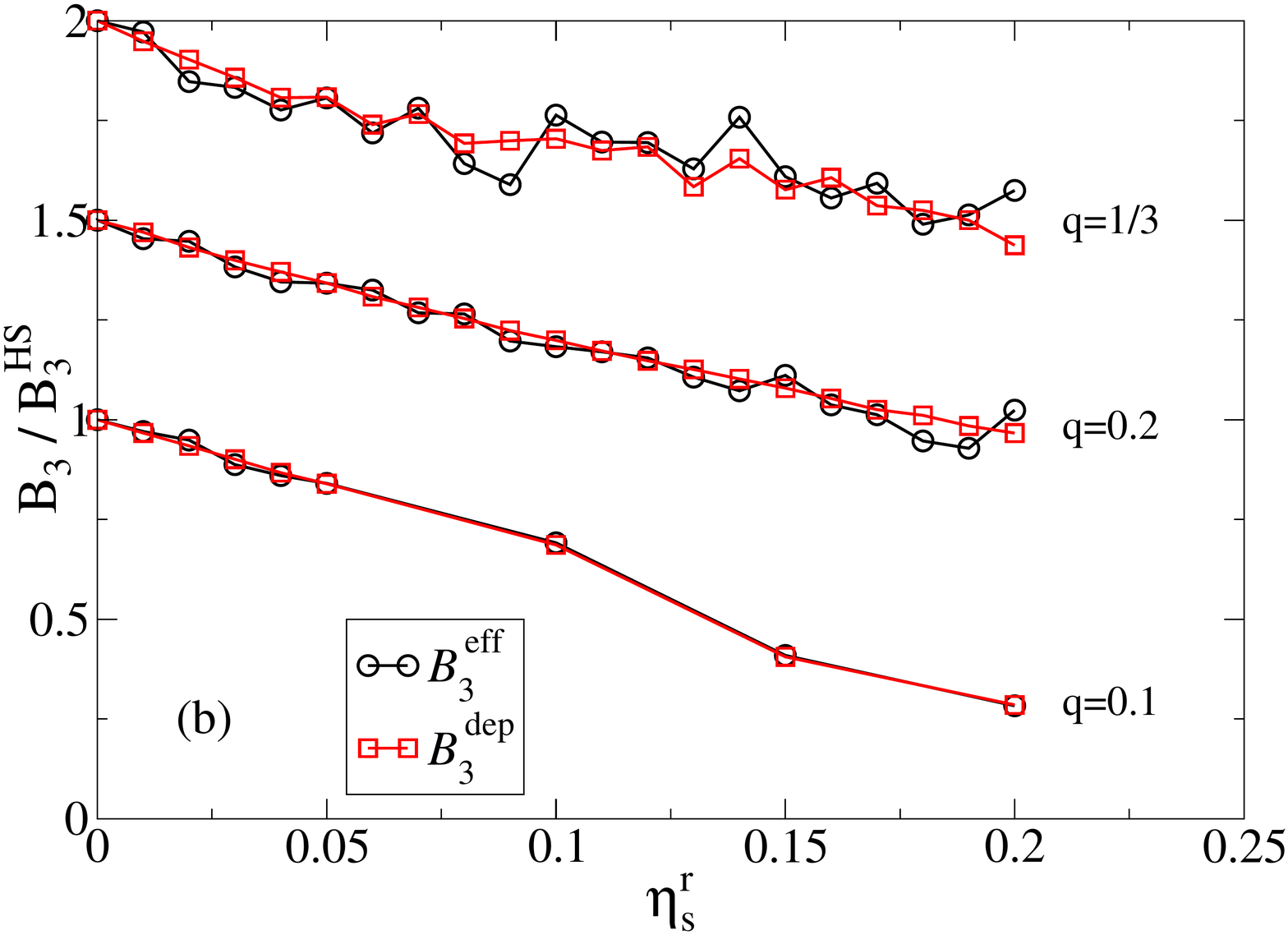}

 \caption{{\bf (a)} Simulation estimates of $P_3(\rmin)$,
   $P_3^{ig}(\rmin)$ and their ratio $g^\prime_3(\rmin)$ as obtained
   in an MC simulation of additive hard spheres for $q=0.2$ and $\eta_s^r=0.1$. The
   volume of the periodic box is $V=(3\sigma)^3$.  {\bf (b)}
   Comparison for size ratio $q=0.1$, $0.2$ and $1/3$ and various
   $\eta_s^r$ of the measured values of $B_3$ arising from the
   simulation of the full two component additive hard sphere model
   [$B_3^{\rm eff}(\eta_s^r)$], and a three particle system
   interacting via the depletion potential [$B_3^{\rm
       dep}(\eta_s^r)$]. The curves for $q=0.2$ and $q=1/3$ have been
   shifted vertically by $0.5$ and $1.0$ respectively.}

 \label{fig:B3hs}
 \end{figure}

Although three-body interactions are always present in principle, our
results shown in fig.~\ref{fig:B3hs}, indicate that within this more
limited range of $\eta_s^r$, they are negligibly small for $q=0.2$ and
$q=0.1$. This finding suggests that for applications at low to
moderate $\eta_s^r$ and small $q$ it is safe to use depletion
potentials for additive hard spheres. Unfortunately, we are unable to
provide indications of the scale of three-body interactions in the
regime of putative phase separation \cite{Ashton:2011kx} which lies
above $\eta_s^r=0.3$. To do so would require a more efficient
simulation algorithm for dealing with mixture of additive hard spheres
than is currently available.

Compared to the AO model the statistical noise on our estimates of
$B_3$ are greater for additive hard spheres, particularly at larger
$q$. There are a number of factors contributing to this extra noise:
The interaction range is longer due to correlations in the solvent (as
shown in fig. \ref{fig:B3hs}(a)), while the overall interaction
strength is weaker for a given $\eta_s^r$. We therefore require a
larger box to access the asymptotic regime and the value of the
finite-size signal $f(V)-1$ will be smaller, leading to some loss of
precision as described in Sec.~\ref{sec:optimise}. This is
particularly apparent at larger $q$ as seen in
fig. \ref{fig:B3hs}(b). The lowest statistical noise is for $q=0.1$
where we were able to use a smaller box, $L=3.5\sigma$, as opposed to
$L=4\sigma$ for the other $q$ values. In addition, because the GCA
algorithm is considerably less efficient for hard sphere solvents, the
amount of statistics we can gather is much less compared to the AO
model for equal computing time.

\section{Application II: Three-body interactions in star polymers}
\label{sec:stars}

As a further application of our method, we have used it to quantify
the scale of three-body interactions in coarse-grained models for star polymers in
implicit solvent. We model each polymer in terms of a core particle
which is functionalised by a number of linear polymer chains each
comprising $n$ monomers. Bonded monomers interact via a FENE
spring~\cite{Kremer:1990rt}, while non-bonded monomers experience a
Lennard-Jones (LJ) potential. Using the Lammps Molecular Dynamics package
\cite{Plimpton:1995ly} we have studied various combination of
functionality and chain length $n$. Our aim was to determine how these
parameters affect the size of the three-body interactions.

Varying functionality and arm chain length leads to overall changes in
the balance of attraction and repulsion between molecules ie. to the
second virial coefficient $B_2$. Accordingly, in order to isolate the
influence of three-body interactions we have, for each combination of
functionality and $n$ studied, measured $B_3$ at a fixed value of
$B_2$.  This was achieved by using histogram reweighting
\cite{Ferrenberg1988} to extrapolate our measured data for
$g^\prime_2(r)$ with respect to temperature such to to precisely
locate that temperature for which $B_2$ matches a prescribed
value. This value was chosen to correspond to a moderately attractive
system, corresponding to a somewhat poor solvent.

\begin{figure}[h]
\includegraphics[type=pdf,ext=.pdf,read=.pdf,width=0.9\columnwidth,clip=true]{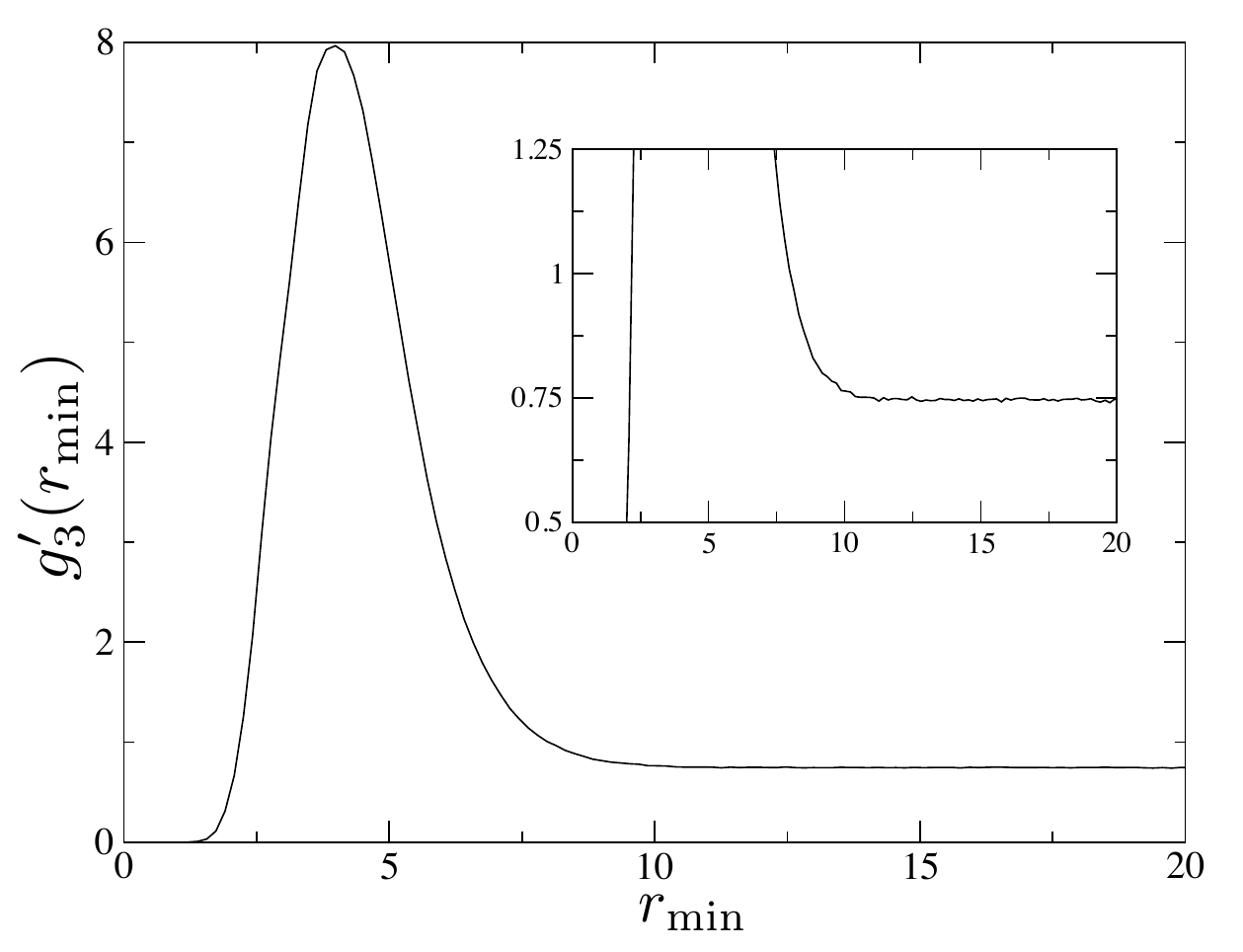}

\caption{Example of the form of $g_3^\prime(r_{\rm min})$ obtained in
  a study of three $7$-armed stars with arm chain length $n=10$. The box volume is $V=(40\sigma)^3$. The inset focuses on the asymptotic region.}
 \label{fig:g3rstar}
 \end{figure}

The procedure for measuring the size of three-body interactions via
virial coefficients is similar to that outlined for the
colloid-polymer mixtures, except that the tagged particles are now
taken to be the set of core atoms. The pair potential is the potential
of mean force (pmf) which is obtained in a simulation of two stars. We
then simulate three particles interacting via this potential to obtain
$B_3^{\rm pmf}$. This we compare with $B_3^{\rm star}$, measured in a
simulation of $N=3$ star polymers. An example plot of $g_3^\prime(r_{\rm
  min})$ is shown in 
Fig.~\ref{fig:g3rstar} measured for a system of three star polymers each having $7$ arms of length $n=10$.

\begin{figure}[h]
\includegraphics[type=pdf,ext=.pdf,read=.pdf,width=0.95\columnwidth,clip=true]{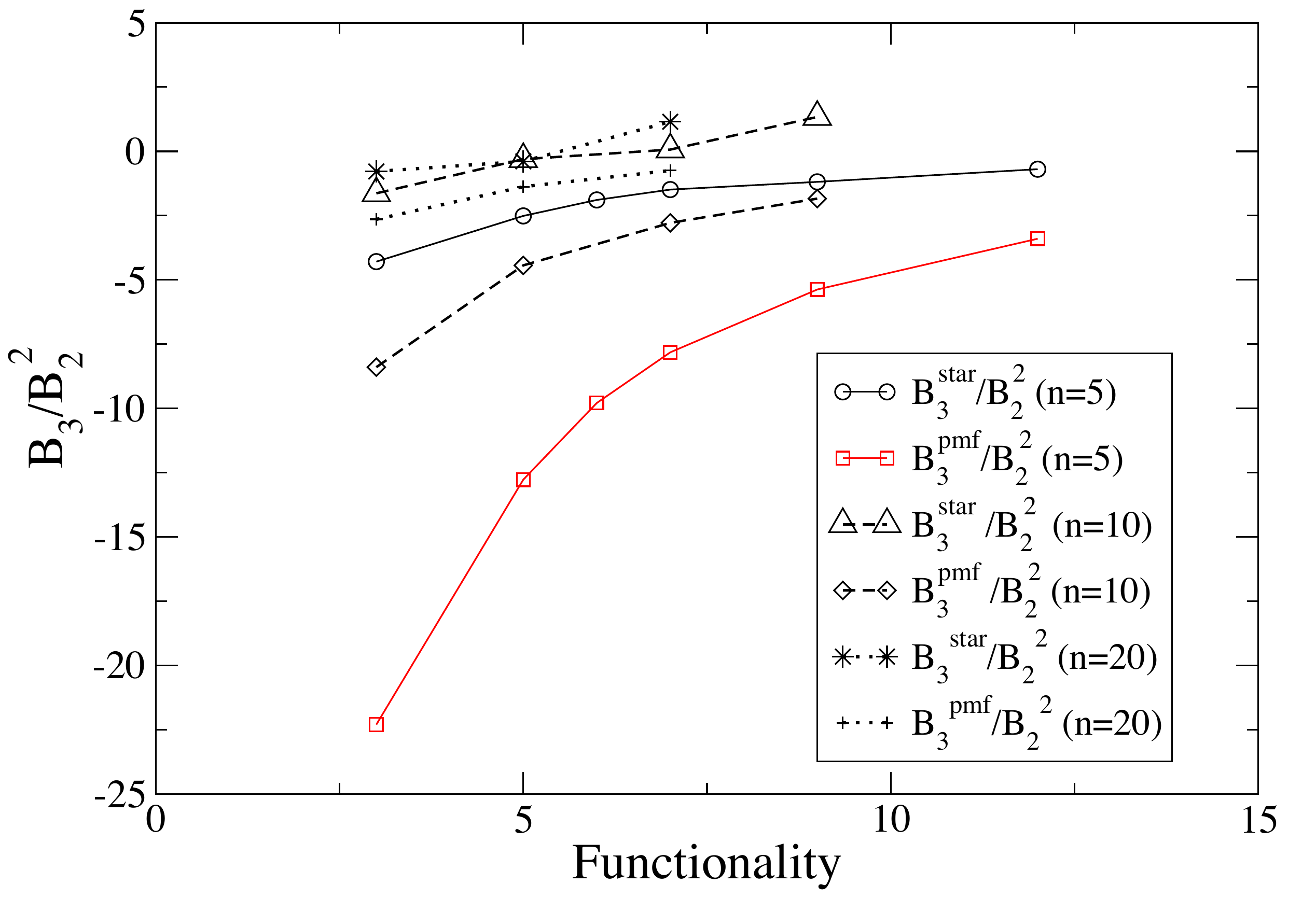}
\caption{Estimates of the dimensionless third virials $B_3^{\rm pmf}/B_2^2$ and $B_3^{\rm star}/B_2^2$ versus  functionality for various chain lengths $n$. Volumes ranged
from $V=(20\sigma)^3$ to $V=(40\sigma)^3$, large enough to
  access the limiting behaviour of $g_3^\prime(r_{\rm min})$. Bonded
  monomers interact via a FENE potential with parameters
  $K=30.0\epsilon/\sigma^2, R_0=1.5\sigma$~\cite{Kremer:1990rt}. The
  LJ potential was truncated and shifted at $r=2.5\sigma$. In all cases
 $T$ is chosen to yield $B_2=-3321\sigma^3$. Errors are comparable with symbol sizes.}
 \label{fig:star_results}
 \end{figure}

Our collected results are shown in Fig.~\ref{fig:star_results} and reveal
considerable discrepancies between $B_3^{\rm pmf}$ and $B_3^{\rm
  star}$, particularly for small values of the functionality and the
arm length. Clearly the disparity is such that one should expect a
quite different equation of state (as well as other thermodynamical
quantities) to arise from the coarse-grained system described by the
pmf compared to the full model. 

The important role of many-body effects in this system arises from the
ability of the polymers to occupy the same volume.  When two polymers
overlap, the local density of the resulting composite particle is much
greater than for a single polymer. Accordingly a third polymer is much
less likely to overlap with the first two due to short ranged
monomeric repulsions. However, this effect is completely neglected in
the pair potential framework for which the degree of attraction is
purely additive. This finding should be relevant to many other types
of polymer-based soft particles, including cluster forming amphiphilic
dendrimers~\cite{Lenz:2012fk}.

\section{Summary}

\label{sec:discuss}

We have proposed a technique for studying three-body interactions in
coarse-grained models for complex fluids.  The method rests on
measurements, for $N=2$ and $N=3$ particles, of the asymptote of a
simple-to-measure structural function $g_N^\prime(r_{\rm min})$.  For
finite simulation box volumes this asymptote yields the ratio of
configurational integrals $Z_1^N/Z_N$, which in turn provides
estimates of virial coefficients.  Comparison of the measured value of
the third virial coefficient of the full system with that of an
approximate coarse-grained representation (involving only pair
potentials) provides information on the scale of three-body
interactions in the effective Hamiltonian. Since the method is based
solely on structural information, it can be utilized with any
simulation scheme capable of generating equilibrium configurations.

Our approach follows the partition function route to virial
coefficients. Consequently it struggles to access high order virial
coefficient for reasons discussed in Sec.~\ref{sec:optimise}. For simple
fluids other methods based on the Mayer function route (such as the
MSM) allow one to reach higher order virial coefficients. However for
calculating low order virials of complex fluids the present method
appears to be a useful tool. This is because it focuses attention on
the configurational statistics of a single atom in the molecule; there is no
need to explicitly integrate over molecular conformations. Being based
on structural information it can also be applied in situations where
the Mayer route fails, namely effective models for mixtures that arise
from tracing out the degrees of freedom of one species.

We have applied the method to calculate the contribution of three-body
interactions to the third virial coefficient of the effective
Hamiltonian in size asymmetrical hard sphere mixtures which serves as
a prototype model for colloid-polymer mixtures or molecular
solutions. Using the geometrical cluster algorithm, we were able to
scan a wide range of size ratios $q$ and small particle reservoir
volume fractions $\eta_s^r$. In the Asakura-Oosawa model at large $q$
and large $\eta_s^r$ three-body effects were found to be substantial;
for example at $q=0.5$ and $\eta_s^r=0.5$, we find $B_3^{\rm
  eff}/B_3^{\rm dep}\approx 3$. However, as expected on decreasing
$q$, the differences diminish rapidly and, for $q=0.154$, were found
to be identically zero within numerical uncertainty for all $\eta_s$.

Compared with the AO model, additive hard spheres present a
significantly greater computational challenge. The range of $\eta_s^r$
accessible to the cluster algorithm is much smaller than the AO model,
being limited to $\eta_s^r<0.2$. Our results indicate that within this
more limited range of $\eta_s^r$, three-body interactions are
negligibly small for $q=0.2$ and $q=0.1$. This finding suggests that
for applications at low to moderate $\eta_s^r$ and small $q$ it is
safe to use depletion potentials for additive hard spheres.

Finally, our study of star polymers modelled as soft effective spheres,
revealed large three-body contributions to the effective potential
which are neglected in the pair potential picture. It was argued that
it is the ability of such molecules to substantially overlap which
leads to the failure of the pair potential approach in such systems.

\acknowledgments

This work was supported by EPSRC grants EP/F047800 and EP/I036192. Computational
results were partly produced on a machine funded by HEFCE's Strategic
Research Infrastructure fund. We thank Bob Evans, Rob Jack, Andrew
Masters and Friederike Schmid for useful discussions.

\appendix
\section{Virial coefficients in terms of configurational integrals}

Here we outline the derivation of the relationships between the virial coefficients,
$B_N$, and the partition functions, following the approach of Hill \cite{Hill:1988ye}.
The starting point is the grand partition function for a monodisperse assembly of particles
\be
\Omega = \sum_{N=1}^\infty \frac{1}{N!}z^{N} Z_N\:.
\ee
Here $z=e^{\mu}$, is the fugacity and $Z_N$ is the $N$-particle configurational integral:

\be
Z_N =  \int_{\br^{N}} \mathrm{d} \br^{N} \exp {(-\beta H(\br^N))}\:.
\ee
The pressure is related to $\Omega$ by
\be
\beta P V = \log \Omega\:,
\ee
and we can take derivatives of the grand partition function to get
\be
\langle N \rangle = \rho V = z \frac{\partial \log \Omega}{\partial z}\:.
\ee
By taking functional derivatives we can expand $\Omega$ as a series in the fugacity:
\be
\frac{\log \Omega}{V} = \beta P = \sum_{j=1}^\infty b_j z^j\:,
\ee
and also the density
\be
\rho = \sum_{j=1}^\infty j b_j z^j\:.
\ee

The coefficients, $b_j$, are the cluster integrals, which are related \cite{Hill:1988ye} to the semi-invariants

\be
j! V b_j = j! \sum_{\bf{n}} \biggl ((-1)^{\sum_{i=1}^j n_i-1} \bigl(  \sum_{i=1}^j n_i -1 \bigr)! \prod_{k=1}^j \biggl [\frac{(Z_k / k!)^{n_k}}{n_k!} \biggr ] \biggr)
\ee
where the first sum is over all sets of positive integers or zero,
$\bf{n}$, such that $\sum_i i n_i=j$.  Thus for $j=1$ the only
possibility is $n_1=1$, while for $j=2$, one has  $n_1 n_2=0 1$ and $2 0$, and for
$j=3$ the possibilities are $n_1 n_2 n_3=0 0 1$, $1 1 0$ and $3 0 0$. The resulting
invariants are

\begin{eqnarray}
1! V b_1 &=& Z_1 \\
2! V b_2 &=& Z_2-Z_1^2 \\
3! V b_3 &=& 2 Z_1^3-3 Z_2 Z_1+Z_3 \\
4! V b_4 &=& -6 Z_1^4+12 Z_2 Z_1^2-4 Z_3 Z_1-3 Z_2^2+Z_4  \\\nonumber
%5! V b_5 &=& 24 Z_1^5-60 Z_2 Z_1^3+20 Z_3 Z_1^2+30 Z_2^2 Z_1-5 Z_4 Z_1-10 Z_2 Z_3+Z_5
%6! V b_6 &=& -120 Z_1^6+360 Z_2 Z_1^4-120 Z_3 Z_1^3-270 Z_2^2 Z_1^2+30 Z_4 Z_1^2+\\ 
%&\ & 120 Z_2 Z_3 Z_1-6 Z_5 Z_1+30 Z_2^3-10 Z_3^2-15 Z_2 Z_4+Z_6 \nonumber
\end{eqnarray}

To obtain virial coefficients, one starts with the virial expansion of the pressure.
\be  
\beta P = \rho + B_2 \rho^2 + B_3 \rho^3 = \rho + \sum_{N=2}^\infty B_N \rho^N\:.
\label{eq:virial}
\ee
Then recall that
\be \label{eq:pex}
 \beta P = \sum_{j=1}^\infty b_j z^j\:,
\ee
and
\be \label{eq:rhoex}
\rho = \sum_{j=1}^\infty j b_j z^j\:.
\ee
Substituting \ref{eq:rhoex} into \ref{eq:virial} and equating
coefficients in $z$ with \ref{eq:pex} yields the virial coefficients
in terms of the cluster integrals:

\begin{eqnarray}
B_1 &=& 1 \\
B_2 &=& -\frac{b_2}{b_1^2} \\
B_3 &=& \frac{4 b_2^2-2 b_1 b_3}{b_1^4} \\
B_4 &=& \frac{-20 b_2^3+18 b_1 b_3 b_2-3 b_1^2 b_4}{b_1^6} \\\nonumber
%B_5 &=& \frac{2 \left(56 b_2^4-72 b_1 b_3 b_2^2+16 b_1^2 b_4 b_2+b_1^2 \left(9 b_3^2-2 b_1 b_5\right)\right)}{b_1^8} \nonumber
%B_6 &=& \frac{1}{b_1^{10}} \biggl [-672 b_2^5+1120 b_1 b_3 b_2^3-280 b_1^2 b_4 b_2^2 \\ 
% & & +5 b_1^2 \left(10 b_1 b_5-63 b_3^2\right) b_2 -5 b_1^3 \left(b_1 b_6-12 b_3 b_4\right) \biggr ] \nonumber
\end{eqnarray}
Finally, substituting for configurational integrals, one finds

\begin{widetext}
\begin{eqnarray}
B_2 &=& \frac{1}{2} V \left(1-\frac{Z_2}{Z_1^2}\right) \\
B_3 &=& \frac{V^2 \left(Z_1^4-3 Z_2 Z_1^2-Z_3 Z_1+3 Z_2^2\right)}{3 Z_1^4} \\
B_4 &=& \frac{V^3 \left(2 Z_1^6-12 Z_2 Z_1^4-8 Z_3 Z_1^3+\left(27 Z_2^2-Z_4\right) Z_1^2+12 Z_2 Z_3 Z_1-20 Z_2^3\right)}{8 Z_1^6} \\\nonumber
%B_5 &=& \frac{V^4 }{30 Z_1^8} \biggl [ 6 Z_1^8-60 Z_2 Z_1^6-60 Z_3 Z_1^5+15 \left(15 Z_2^2-Z_4\right) Z_1^4 
%& & +\left(200 Z_2 Z_3-Z_5\right) Z_1^3- 5 \left(72 Z_2^3-4 Z_4  Z_2-3 Z_3^2\right) Z_1^2 \nonumber \\
% & & -180 Z_2^2 Z_3 Z_1+210 Z_2^4\biggr ] \nonumber \\
% B_6 &=& \frac{V^5}{144 Z_1^{10}}   \biggl [  24 Z_1^{10}-360 Z_2 Z_1^8-480 Z_3 Z_1^7+180 \left(11 Z_2^2-Z_4\right) Z_1^6+ \\ 
% & & 24 \left(110 Z_2 Z_3-Z_5\right) Z_1^5-\left(5100 Z_2^3-525 Z_4 Z_2-400 Z_3^2+Z_6\right) Z_1^4 \nonumber \\
% & & +30 \left(-170 Z_3 Z_2^2+Z_5 Z_2+2 Z_3 Z_4\right) Z_1^3+210 Z_2 \left(30 Z_2^3-2 Z_4 Z_2-3 Z_3^2\right) Z_1^2 \nonumber \\
% & & +3360 Z_2^3 Z_3 Z_1-3024 Z_2^5 \biggr ] \nonumber
\end{eqnarray}
\end{widetext}

\end{document}